\documentclass[%
 reprint,
 amsmath,amssymb,
 aps,
 pra,
 showkeys,
 floatfix,
]{revtex4-2}

\usepackage{multirow} 
\usepackage[utf8]{inputenc}
\usepackage[T1]{fontenc}
\usepackage{graphicx}
\usepackage{dcolumn}
\usepackage{bm}
\usepackage{booktabs}
\usepackage{array}
\usepackage[colorlinks=true, linkcolor=blue, citecolor=blue, urlcolor=blue]{hyperref}
\usepackage[capitalise]{cleveref}
\usepackage[all]{hypcap}
\usepackage{booktabs} 
\usepackage{amsmath}
\usepackage{amssymb}
\usepackage{mathtools}
\usepackage{mathrsfs}
\usepackage{amsthm}
\usepackage{amssymb}
\usepackage{url}
\usepackage{makecell}
\usepackage{xcolor,soul}
\sethlcolor{yellow}
\usepackage[ruled,vlined,linesnumbered]{algorithm2e}
\soulregister\cite7
\usepackage{float}

\begin{document}

\preprint{APS/123-QED}

\title{Multi-tasking through quantum annealing}

\author{Jargalsaikhan Artag}
\author{Koki Awaya}
\author{Takumi Kanezashi}
\author{Daisuke Tsukayama}
\author{Moe Shimada}
\author{Jun-ichi Shirakashi}
\email{shrakash@cc.tuat.ac.jp}
\affiliation{Department of Electrical Engineering and Computer Science, Tokyo University of Agriculture and Technology, Koganei, Tokyo, Japan.}

\date{\today}

\begin{abstract}
Quantum annealing approximately solves combinatorial optimization problems by leveraging the principles of adiabatic quantum systems. In this approach, the system's Hamiltonian evolves from an initial general state to a problem-specific state. This study introduces multi-tasking quantum annealing (MTQA), a method that enables the parallel processing of multiple optimization problems by embedding them into spatially distinct regions on quantum hardware. MTQA is evaluated using two NP-hard problems: the minimum vertex cover problem (MVCP) and the graph partitioning problem (GPP). This parallel approach optimizes quantum resource utilization by concurrently utilizing idle qubits. The findings demonstrate that MTQA achieves a solution quality comparable to single-problem quantum annealing and classical simulated annealing (SA), while notably reducing the time-to-solution (TTS) metrics. Eigenspectrum analysis further theoretically supports the hypothesis that parallel embedding preserves quantum coherence and does not increase computational complexity by efficiently utilizing available quantum hardware (e.g., qubits and couplers). MTQA enables efficient multitasking in quantum annealing, optimizing hardware utilization and improving throughput for concurrent tasks and demonstrating performance for problems up to 100 nodes in real-world applications.
\end{abstract}


\maketitle

\section{Introduction}

Quantum annealing has emerged as a promising approach for solving complex combinatorial optimization problems by leveraging the quantum mechanical properties to explore extensive solution spaces\cite{Farhi2000,Morita2008}. This technique is grounded in the principle of adiabatic quantum computation, where solutions are encoded in the ground state of a time-dependent quantum Hamiltonian\cite{Farhi2000,Morita2008,AlbashLidar2018}. Initially, the system is prepared in the ground state of a simple initial Hamiltonian and then evolves slowly into the ground state of a problem-specific Hamiltonian that represents the desired optimization problem\cite{KadowakiNishimori1998}. The Hamiltonian that governs the quantum annealing process contains two main components: the initial Hamiltonian and the problem Hamiltonian. This can be expressed as follows:

\begin{equation}
    \begin{split}
        H(s) ={}& -\frac{A(s)}{2}\left(\sum_i \sigma_i^x\right) \\
        & + \frac{B(s)}{2}\left(\sum_i h_i\sigma_i^z + \sum_{i>j} J_{i,j}\sigma_i^z\sigma_j^z\right),
    \end{split}
\end{equation}

where the Pauli-X and Pauli-Z operators acting on the $i$-th qubit are denoted as $\sigma_i^x$ and $\sigma_i^z$ respectively. The time-dependent coefficients $A(s)$ and $B(s)$ control the evolution of the Hamiltonian from an initial, non-interacting state toward the state that encodes the optimization problem. Initially, $A(s)$ dominates, placing the system in a superposition of all possible states, which is essential for exploring the solution space. As the process progresses, $B(s)$ takes precedence, navigating the system toward a problem-specific configuration that represents the optimal solution\cite{Hauke2020,DasChakrabarti2005}.

Quantum annealing has been implemented on platforms such as D-Wave systems, which employ superconducting circuits to execute the transverse-field Ising model, as shown in equation (1)\cite{Johnson2011}. This model is specifically suited for solving problems formulated as quadratic unconstrained binary optimization (QUBO)\cite{Boixo2014}. Various quantum optimization algorithms are being explored across different hardware paradigms. For example, the Digitized-Counter diabatic Quantum Approximate Optimization Algorithm (DC-QAOA)\cite{Chandarana2022} has shown promise on gate-model quantum computers. However, this study focuses on maximizing the utility of existing and future quantum annealing platforms.

A significant challenge in the implementation of quantum annealing is the efficient utilization of limited quantum hardware resources. Although current quantum annealers demonstrate potential for solving certain combinatorial optimization problems, they face constraints related to the number of available qubits and their connectivity\cite{Boixo2014}. As Quantum Processing Unit (QPU) size increases (e.g., 5000+ qubits), embedding single instances of small-to-medium scale problems often leaves most of the processor idle. To address these constraints and improve throughput in multi-user or cloud environments, spatial multiplexing techniques such as  Parallel Quantum Annealing (PQA) were  developed\cite{Pelofske2022,Artag2023PQA,Artag2024MTQA,Pelofske2023Clique,Pelofske2024QAOA,Huang2023} to optimize hardware utilization by embedding multiple independent problem instances simultaneously. Although PQA demonstrates the general feasibility of parallel processing, standard baseline implementations often employ global QUBO scaling or a unified approach to chain strength calculation for simplicity. This can be suboptimal for heterogeneous tasks where concurrently processed problems may have differing intrinsic energy scales, structural complexities, or optimal annealing parameters\cite{Pelofske2022,Artag2023PQA,Artag2024MTQA,Pelofske2023Clique,Pelofske2024QAOA,Huang2023}. This can lead to compromised solution accuracy for some instances within a parallel batch, which is a challenge highlighted by the comparative results in this work. To address these limitations and enable more robust parallel performance, we introduce multi-tasking quantum annealing (MTQA). MTQA enhances parallel embedding by preserving the individual characteristics of each co-scheduled problem, offering significantly improved parameter control, problem isolation, and consequently, more reliable solution quality across diverse tasks.

Building on the foundation of PQA, this study introduces MTQA. Unlike classical computing, where multi-tasking involves parallel task execution to enhance efficiency, MTQA enables parallel processing of multiple optimization problems by exploiting the spatial separation of problem embeddings on quantum hardware, where physically distinct qubit regions can be annealed simultaneously without the need for inter-problem communication or coordination during the annealing process\cite{Huang2023,Blake2009,HennessyPatterson2011}. MTQA ensures high solution quality through several key technical innovations designed for precise problem-specific parameter control. First, MTQA offers an optional isolation-layer strategy, which is buffer zones of unused qubits between distinct embedded problem instances. This is designed to mitigate potential residual spurious couplings and hardware interference, thereby enhancing effective problem separability on the QPU. Second, MTQA treats each embedded problem instance independently and calculates the chain strengths based on the unique characteristics of each embedding. This allows for the accurate handling of identical problems that might have different chain lengths owing to variations in hardware. Third, MTQA performs scaling of embedded QUBOs, where each instance is individually scaled to fit hardware constraints before combination. This prevents problems with naturally smaller energy scales (e.g., those with fewer variables or smaller coefficient magnitudes or different problem types) from being numerically overshadowed or poorly represented when co-scheduled with problems that have much larger intrinsic energy scales, which is a critical factor for robustly handling heterogeneous workloads. Finally, in the solution-extraction phase, MTQA unembeds each problem instance independently. This allows for the application of different unembedding strategies for each problem instance: majority vote, weighted random, or minimum energy. In this study, we focus on the majority vote strategy to analyze the impact of broken qubit chains on solution quality across different problem types.

Although previous work on parallel problem solving using annealers has suggested a risk of reduced solution accuracy compared to single-instance runs\cite{Pelofske2022}, our experimental results demonstrate the essential advantages of MTQA. This risk arises primarily from two factors: (1) potential spurious interactions between adjacent qubit regions when problems are densely embedded without isolation and (2) suboptimal global parameterization (chain strengths and QUBO scaling) that fails to account for the unique characteristics of each problem instance. Our method consistently achieves solution quality equivalent to traditional single-problem quantum annealing across the problems tested and often outperforms in terms of time-to-solution (TTS). Notably, as detailed in the Results section, MTQA significantly outperforms PQA, especially for problem instances of the minimum vertex cover problem (MVCP), where per-instance parameters are key to success. This robust performance in a multi-tasking context is particularly significant given the probabilistic and heuristic nature of quantum annealing, where solution quality can vary across runs and is critically dependent on parameter settings. By leveraging the parallel embedding strategy of MTQA, we can perform multiple samplings of the same problem simultaneously, effectively increasing the probability of finding optimal solutions. The combination of enhanced sampling capabilities and precise problem-specific parameter controls enables a more thorough exploration of the solution space without compromising computational efficiency. This approach not only maximizes the utilization of quantum hardware resources but also improves the efficiency of quantum annealing systems by increasing sampling statistics and enhancing solution diversity.

In this study, we explored the practicality and advantages of MTQA by focusing on two NP-hard problems: MVCP\cite{Karp2010,Lucas2014} and the graph partitioning problem (GPP)\cite{Lucas2014,UshijimaMwesigwa2017,FuAnderson1986,MezardMontanari2009}. While the fundamental concept of spatial multiplexing is established, our implementation introduces a systematic framework to resolve the performance degradation commonly observed when concurrently solving heterogeneous tasks. Our key technical contributions include:

\begin{itemize}
\item A systematic, per-instance parameterization framework for chain strength calculation that accounts for the unique embedding characteristics and type of each concurrent problem.
\item An independent QUBO scaling technique applied to each problem before parallel execution, ensuring appropriate Hamiltonian representation and preventing numerical overshadowing among tasks with differing energy scales.
\item An eigenspectrum analysis that provides rigorous theoretical backing for why global parameterization fails for heterogeneous problem sets like MVCP, demonstrating how our per-instance approach prevents spectral gap collapse.
\end{itemize}

The remainder of this paper is organized as follows. In the Results section, we first present our parallel embedding strategy and analyze its impact on chain length distribution. We then examine the performance of MTQA through metrics such as ground-state probability and time-to-solution, comparing these results with those of traditional quantum annealing and classical solvers, such as IBM ILOG CPLEX\cite{Nickel2022} and Gurobi\cite{Bixby2007}. Additionally, we provide an eigenspectrum analysis to examine the quantum mechanical behavior of the system. The Discussion section analyzes the implications of our findings and explores potential applications and future research directions. Finally, the Methods section details the technical implementation and experimental setup.

This work represents a significant step toward making quantum annealing more practical for real-world applications by enabling efficient multitasking without compromising solution quality. Our findings demonstrate that MTQA can effectively utilize quantum resources while simultaneously solving diverse optimization problems, thereby opening new possibilities for quantum computing applications.

\section{Results}

\subsection{Parallel embedding analysis}

\begin{figure}[t]
\centering
\includegraphics[width=\columnwidth]{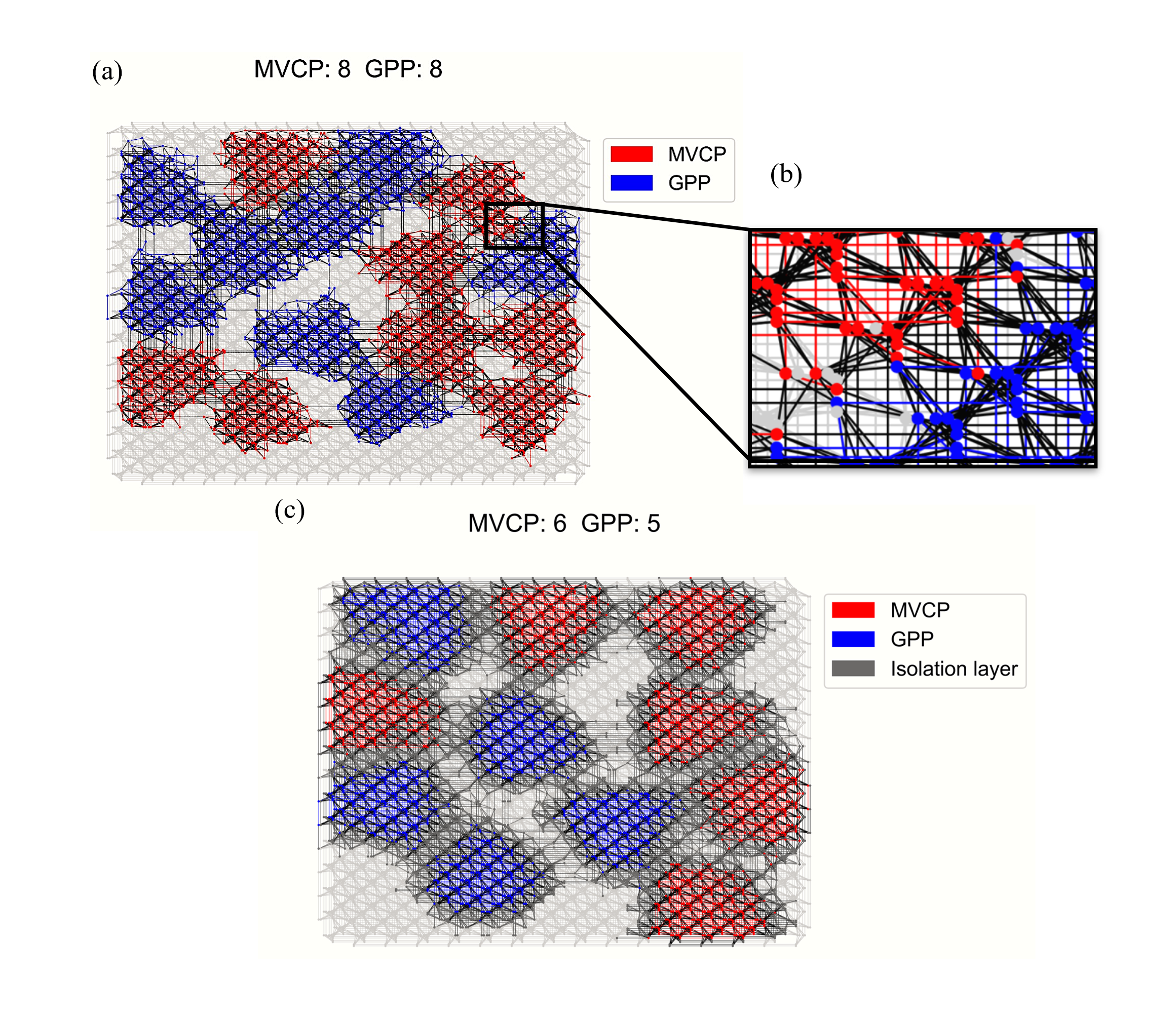}
\caption{Parallel embedding in D-Wave Advantage 6.4 hardware (Pegasus topology). (a) Dense parallel embedding without isolation, where problems are tightly embedded on hardware. MVCP (eight instances, shown in red) and GPP (eight instances, shown in blue) problems embedded into hardware, demonstrating maximum hardware utilization. (b) Magnified view of the region from (a), highlighting congested qubit environment where adjacent qubits support different tasks. (c) Parallel embedding with neighbor isolation strategy (gray regions represent isolation layers), which separates different problem instances on hardware. We embedded MVCP (six instances, shown in red) and GPP (five instances, shown in blue) with this method.}
\label{fig:fig1}
\end{figure}

The parallel embedding of multiple problem instances on a D-Wave quantum annealer\cite{DWaveQPU2024,Boothby2020} is fundamental to the operation of MTQA. This technique enables processing of distinct problems while optimizing quantum resource utilization. Our investigation, conducted on a D-Wave Advantage 6.4 quantum annealer, identified two distinct approaches for implementing MTQA: dense embedding without isolation and embedding with isolation layers. Figure~\ref{fig:fig1} shows these approaches using problem instances of MVCP and GPP, each with 40 nodes and a 0.9 edge density probability.

The dense embedding approach [Fig.~\ref{fig:fig1}(a)] maximizes hardware utilization by embedding eight MVCP and eight GPP instances simultaneously. A magnified view [Fig.~\ref{fig:fig1}(b)] reveals a complicated qubit connectivity pattern in which adjacent qubits are shared among different problem instances. This configuration often leads to longer chain lengths because qubits adjacent to one problem instance may already be allocated to neighboring problems. Although this approach achieves the highest number of parallel embeddings, it creates a congested environment that increases the risk of hardware interference and spurious couplings between adjacent qubits, which could potentially affect solution quality during the annealing process.

With the isolation strategy [Fig.~\ref{fig:fig1}(c)], we introduced buffer zones, shown in gray, between problem instances, ensuring that no two problems are directly adjacent, thus preventing interference and improving problem isolation. As a trade-off, the number of embedded problems is reduced to six MVCP and five GPP instances. This isolation layer was introduced to buffer against known hardware noise phenomena, specifically spurious couplings and integrated control errors (ICE) \cite{Harris2009,Harris2010,Vuffray2022,Cenk2025} that can arise between neighboring qubits in analog quantum annealers. As detailed in Section 2.3, while both isolated and non-isolated strategies perform similarly for smaller problems, our empirical data demonstrates that the isolation strategy provides a measurable advantage in solution quality for larger, densely connected problems (e.g., yielding the best heuristic performance for GPP at $n=90$).

\begin{figure}[t]
\centering
\includegraphics[width=\columnwidth]{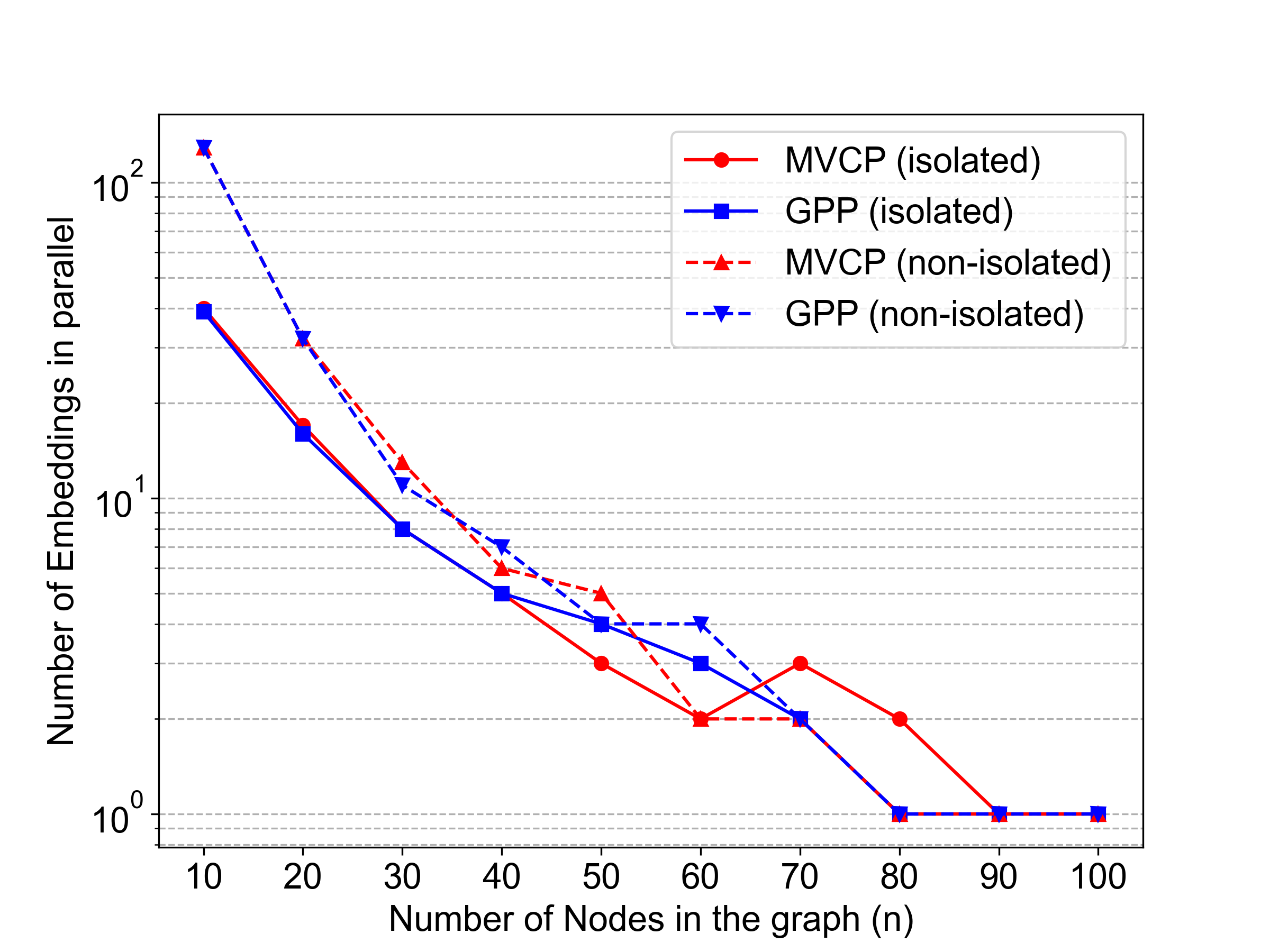}
\caption{Number of parallel embeddings with log scale as a function of graph size. Plot demonstrates how embedding capacity decreases with increasing problem size, with both isolated (solid lines) and non-isolated (dashed lines) strategies shown for MVCP (red) and GPP (blue). Both strategies show exponential decrease in embedding capacity as problem size increases, with convergence for larger problems.}
\label{fig:fig2}
\end{figure}

The relationship between problem size and embedding capacity is illustrated in Fig.~\ref{fig:fig2}. For smaller problems ($n < 30$, where $n$ denotes the number of nodes in the graph), we observed a significant parallel embedding capability, with up to 130 instances for MVCP and 128 instances for GPP in the non-isolated configuration. However, as the problem size increases, the number of feasible parallel embeddings decreases exponentially due to the growing chain lengths and more stringent connectivity requirements. Both the isolated and non-isolated strategies showed similar trends, with the non-isolated approach consistently supporting a higher number of parallel embeddings across all problem sizes. This difference becomes less pronounced as the problem size increases beyond $n = 50$, where both approaches converge to similar embedding capacities owing to the inherent limitations of hardware connectivity.

We note that the stochastic nature of the minorminer embedding algorithm can occasionally result in the isolated strategy supporting more embeddings than the non-isolated one for specific problem sizes (e.g., $n = 70, 80$ MVCP), highlighting the inherent variability in heuristic embedding searches.

\subsection{Chain length analysis}

\begin{figure}[t]
\centering
\includegraphics[width=\columnwidth]{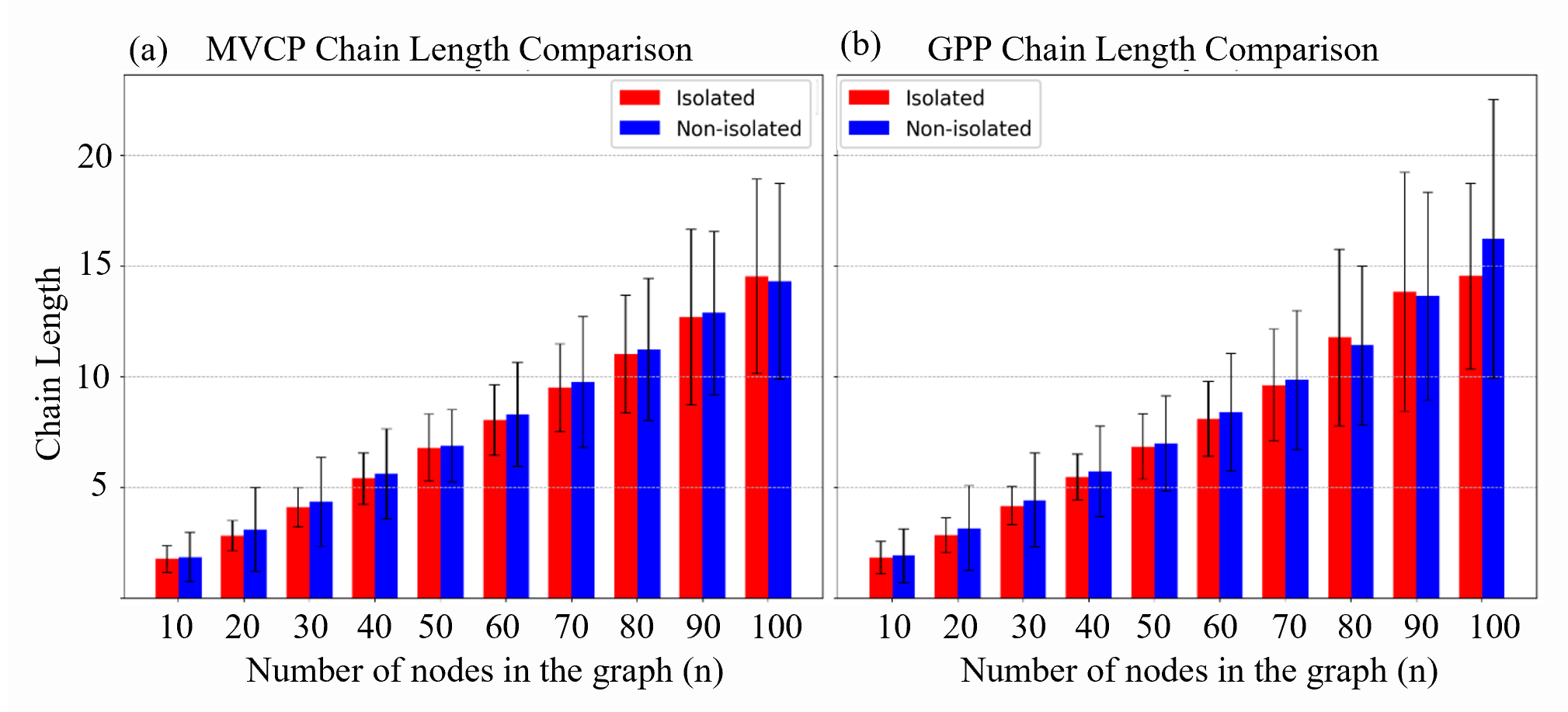}
\caption{Chain length variations as a function of graph size. (a) MVCP chain length comparison showing average chain lengths (bars) and standard deviations (error bars) from 100 parallel embeddings for both isolated (red) and non-isolated (blue) strategies. (b) Similar analysis for GPP, demonstrating consistently longer chains and larger standard deviations, particularly for problems with more than 70 nodes.}
\label{fig:fig3}
\end{figure}

The chain length, which represents the number of physical qubits mapped to a single logical qubit, is influenced by the size and connectivity demands of the problem\cite{DWaveQPU2024,Boothby2020,Cai2014,Djidjev2023}. As the problem size and connectivity requirements increase, so does the chain length, which in turn affects the number of problem instances that can be embedded\cite{Choi2008,Choi2011,Venturelli2015,King2018,Vinci2015}. Figure~\ref{fig:fig3} presents our analysis of chain lengths based on extensive experiments using problems of varying sizes.

We performed 100 parallel embedding trials for each problem size owing to the randomness inherent in the minorminer's \texttt{find\_embedding} algorithm\cite{Cai2014,Djidjev2023}. Figure~\ref{fig:fig3} presents our chain length analysis through two key metrics: the average chain length (shown as bars) and standard deviation (shown as error bars) from the 100 trials. For MVCP problems [Fig.~\ref{fig:fig3}(a)], the average chain length increased steadily with graph size, starting at approximately two physical qubits for 10-node problems and reaching approximately 14 physical qubits for 100-node problems. The standard deviation also increased with the problem size, indicating greater variability in embedding configurations as the problems grew larger. GPP problems [Fig.~\ref{fig:fig3}(b)] demonstrate similar scaling behavior, but GPP consistently required longer chains due to its higher connectivity demands. The average chain length ranged from approximately 2 physical qubits for 10-node problems to approximately 16 physical qubits for 100-node problems. The standard deviation was notably larger for GPP than for MVCP, particularly for problems exceeding 70 nodes, suggesting that finding optimal embeddings becomes increasingly challenging for larger, highly connected problems. When comparing the isolated and non-isolated embedding strategies, the difference in the average chain lengths was relatively small. However, the isolated strategy (shown in red) tended to produce slightly shorter chains for smaller problems ($n < 80$), likely due to reduced competition for nearby qubits. This advantage diminished for larger problems, where both strategies converged to similar chain lengths because of the intrinsic connectivity requirements of the problems.

\subsection{Solution quality}

In this study, we evaluated the efficiency of MTQA by analyzing its performance on two NP-hard problems: MVCP and GPP. Although Quantum Annealing (QA) effectively solved MVCP across varying graph sizes, it encountered difficulties with GPP when the graph size exceeded 20 nodes. We compared MTQA (both isolated and non-isolated approaches) against standard single-instance QA and PQA implementations that employ global parameter settings. While classical heuristic methods such as Simulated Annealing (SA) can potentially achieve enhanced performance through extensive parameter tuning (e.g., sweeping over initial temperatures and cooling schedules), we utilized standard, unoptimized SA parameters in this study. This ensures a balanced, out-of-the-box comparison with our quantum annealing framework and avoids the additional computational time overhead associated with classical parameter sweeps. Our analysis is based on the following four key metrics: ground-state probability (GSP), TTS for MVCP, solution energy distribution, and number of cut edges (cut the edges connecting vertices in different subsets, which we aim to minimize) for GPP.

\subsection{Ground-state probability (GSP)}

For parallel-embedded problems, the GSP is calculated as the average probability of finding optimal solutions across multiple embedded instances\cite{Pelofske2022,Artag2023PQA}. GSP is calculated as follows:

\begin{equation}
P_{\mathrm{avg}} = \frac{1}{N_{\mathrm{MVCP}}}\sum_i^{N_{\mathrm{MVCP}}} p_i,
\end{equation}

where $p_i$ is the probability of finding the optimal solution for the $i$-th problem from $N_{\mathrm{MVCP}}$ parallel instances.

\begin{figure}[t]
\centering
\includegraphics[width=\columnwidth]{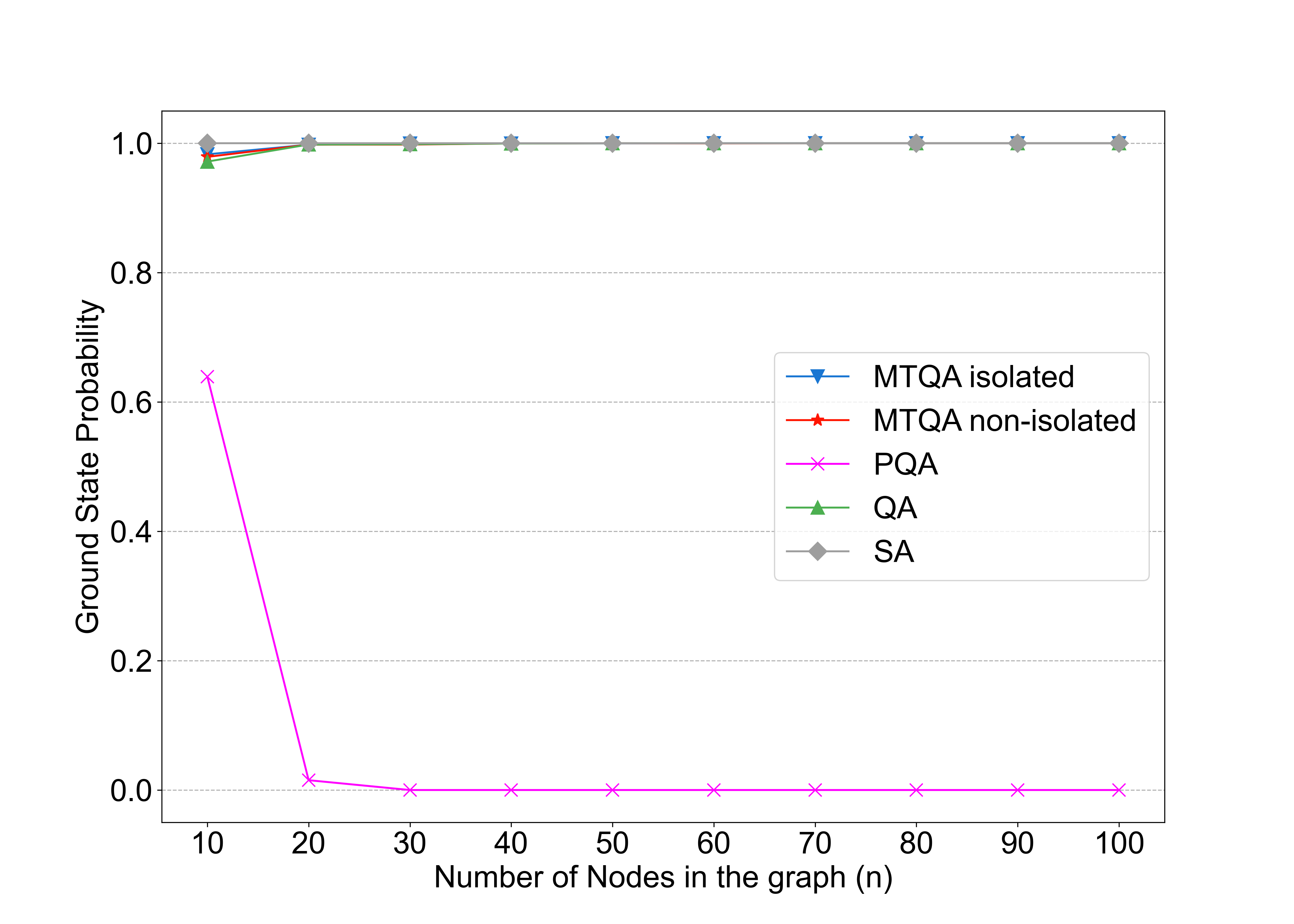}
\caption{Ground-state probability (GSP) analysis comparing quantum and classical approaches. Plot demonstrates GSP values as problem size increases from 10 to 100 nodes. MTQA isolated (blue) and non-isolated (red) implementations compared to PQA (pink), traditional QA (green), and the classical Simulated Annealing (SA) heuristic (gray). MTQA, QA, and SA maintain high GSP ($>0.9$) across most problem sizes. In contrast, PQA fails for heterogeneous MVCP instances, with GSP dropping to near zero for $n \ge 30$.}
\label{fig:fig4}
\end{figure}

Figure~\ref{fig:fig4} shows the GSP results for MVCP. Both MTQA implementations and standard QA demonstrate robust performance, maintaining high GSP values across most problem sizes up to $n = 100$, similar to the classical simulated annealing (SA) benchmark. In contrast, PQA with global parameter settings shows significant performance degradation for MVCP. Although PQA achieves a GSP of 0.64 for $n = 10$, its GSP decreases to approximately 0.01 for $n = 20$ and effectively to zero for all $n \ge 30$. This indicates a failure of the global PQA approach to find optimal solutions for heterogeneous MVCP instances as the problem size increases. The consistent high performance of MTQA compared to PQA highlights the benefit of the per-instance parameter control of MTQA.

\subsection{Time-to-solution (TTS)}

The TTS metric is a key performance indicator for evaluating the efficiency of a quantum annealer in finding the optimal solution to a problem\cite{Zielewski2022,King2015TimeToTarget,Hamerly2019}. It estimates the expected time required to identify the ground state (optimal solution) with a certain probability. Mathematically, TTS is expressed as the ratio of the time taken for one run $t_{\mathrm{run}}$ to the probability of finding the optimal solution\cite{Pelofske2022,Artag2023PQA}. TTS can be expressed as follows:

\begin{equation}
\mathrm{TTS} = t_{\mathrm{run}}\,\frac{\log(1-P_{\mathrm{success}})}{\log(1-P_{\mathrm{avg}})},
\end{equation}

where $P_{\mathrm{success}}$ is the desired success probability, typically set to 0.99, and $P_{\mathrm{avg}}$ is the average success probability across parallel instances. This metric facilitates comparisons of the efficiency of different algorithms and hardware configurations in solving optimization problems. For the MTQA case, TTS is adapted to account for running multiple problems in parallel. Given $t_{\mathrm{run}}$ as the time required for one run and each iteration, we must consider that multiple problems are embedded and solved in parallel. TTS for the same problem embedded multiple times in parallel is calculated as follows:

\begin{equation}
t_{\mathrm{run}} = \frac{1}{A}\left(\frac{T_{\mathrm{QPU}}}{N_{\mathrm{MVCP}}+N_{\mathrm{GPP}}} + \sum_i^{N_{\mathrm{MVCP}}} U_i\right),
\end{equation}

where $A$ is the number of samplings (set to 2500 in our experiments), $T_{\mathrm{QPU}}$ is the D-Wave QPU access time, $U_i$ is the CPU processing time for unembedding solutions for the $i$-th instance of MVCP, and $N_{\mathrm{MVCP}}$ and $N_{\mathrm{GPP}}$ are the number of each problem instance embedded in parallel. Equation (4) accounts for the parallel solving of multiple problems and incorporates both quantum processing and classical post-processing times, providing a complete evaluation of MTQA performance.

\begin{figure}[t]
\centering
\includegraphics[width=\columnwidth]{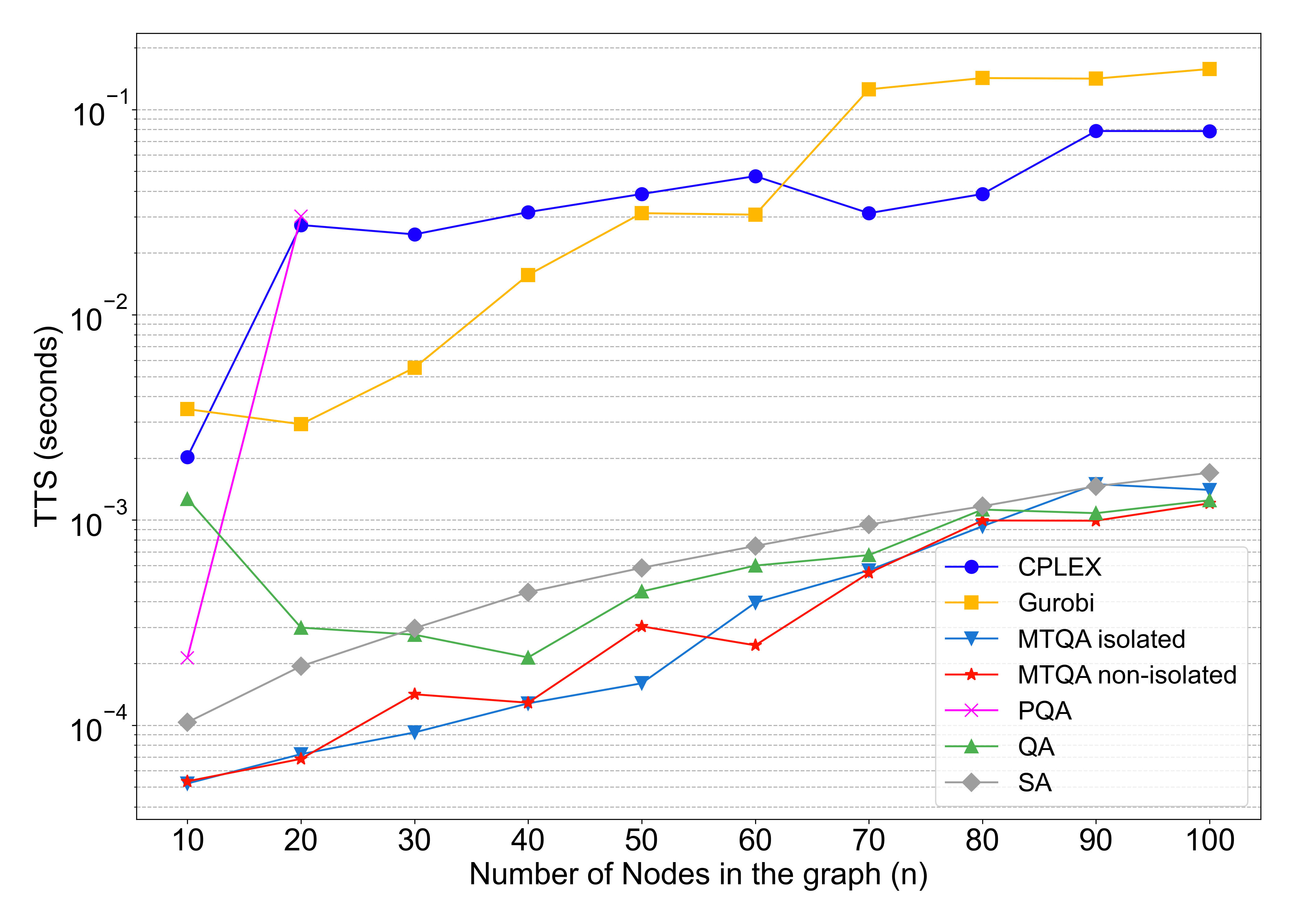}
\caption{Time-to-solution (TTS) comparison across different approaches as a function of graph size. Plot shows TTS (log scale) for problems ranging from 10 to 100 nodes. MTQA approaches (both isolated and non-isolated) achieve lower TTS compared to standard QA and significantly outperform exact solvers in solution generation time (though exact solvers provide optimality guarantees). Simulated Annealing (SA) serves as a classical heuristic benchmark. Note that the PQA line for MVCP terminates at $n = 30$ because its success probability drops to zero, rendering TTS undefined.}
\label{fig:fig5}
\end{figure}

The TTS metrics for MVCP, comparing the various approaches, are shown in Fig.~\ref{fig:fig5}. To ensure a comprehensive evaluation, we benchmarked MTQA against standard QA, PQA, classical heuristic SA, and exact solvers (CPLEX and Gurobi). Both MTQA implementations demonstrate a distinct speed advantage. For problem sizes up to $n = 50$, MTQA achieves lower (better) TTS values than both the standard QA and SA heuristic. Although SA reliably finds the ground state, MTQA finds it faster in this regime. The exact solvers (CPLEX and Gurobi) show the highest TTS values, several orders of magnitude greater than those of the quantum approaches, serving as an optimality baseline rather than a speed benchmark. It should be noted that exact solvers invest significant computational effort in proving optimality, whereas heuristic methods (MTQA, PQA, QA, and SA) focus solely on finding good solutions without optimality guarantees. This fundamental difference in algorithmic objectives makes direct TTS comparison meaningful primarily for understanding the time scale differences rather than declaring superiority of one approach over another. Critically, the PQA line terminates at $n = 30$. This is because global parameterization in PQA causes the success probability to drop to zero for larger MVCP instances, rendering the TTS metric undefined. This contrasts with MTQA, which maintained robust performance across the full range of problem sizes.

\subsection{Number of cut edges in GPP}

Table~\ref{tab:gpp} presents a comprehensive comparison of the number of cut edges obtained using various solving approaches (lower is better). Classical solvers (CPLEX and Gurobi) consistently find optimal solutions, serving as our baseline. For small graphs ($n \le 20$), all quantum methods (MTQA isolated, MTQA non-isolated, QA, and PQA) achieve optimal or very near-optimal solutions. As the problem size increases, all quantum approaches show a slight divergence from the optimal, which is expected for heuristic methods. PQA provides solution quality (cut edges) comparable to both MTQA approaches and QA across all problem sizes. Thus, for GPP solution quality in terms of cut edges, MTQA and PQA perform similarly to standard QA, all providing good heuristic solutions.

\begin{table}[t]
\caption{GPP solution quality measured by cut edges. Best-performing heuristic results for each graph size are marked in bold.}
\label{tab:gpp}
\begin{ruledtabular}
\begin{tabular*}{\columnwidth}{@{\extracolsep{\fill}}cccccccc}
$n$ & CPLEX & Gurobi & SA & \shortstack{MTQA\\isolated} & \shortstack{MTQA\\non-isolated} & PQA & QA \\
\hline
10 & 22 & 22 & \textbf{22} & \textbf{22} & \textbf{22} & \textbf{22} & \textbf{22} \\
20 & 80 & 80 & \textbf{80} & \textbf{80} & \textbf{80} & \textbf{80} & \textbf{80} \\
30 & 187 & 187 & \textbf{187} & 191 & 190 & 189 & 192 \\
40 & 331 & 331 & \textbf{331} & 338 & 338 & 333 & 341 \\
50 & 528 & 528 & \textbf{530} & 542 & 540 & 541 & 541 \\
60 & 767 & 767 & \textbf{772} & 785 & 789 & 779 & 787 \\
70 & 1063 & 1063 & \textbf{1070} & 1082 & 1087 & 1082 & 1085 \\
80 & 1352 & 1352 & \textbf{1369} & 1394 & 1389 & 1379 & 1390 \\
90 & 1752 & 1752 & 1773 & \textbf{1771} & 1791 & 1787 & 1783 \\
100 & 2152 & 2152 & \textbf{2179} & 2195 & 2195 & 2190 & 2191 \\
\end{tabular*}
\end{ruledtabular}
\end{table}

\subsection{Solution energy distribution}

\begin{figure*}[t]
\centering
\includegraphics[width=0.96\textwidth]{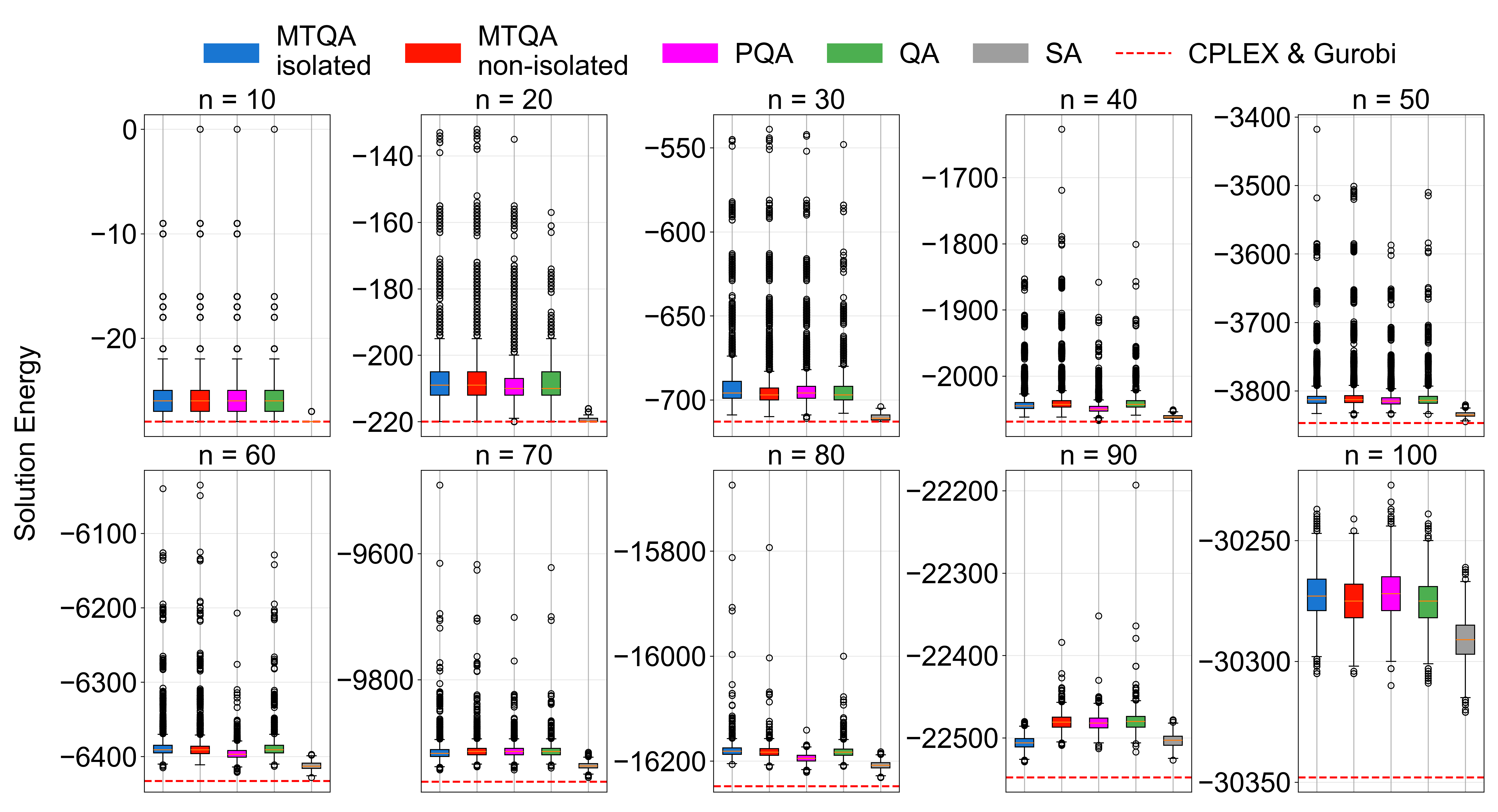}
\caption{GPP solution energy distribution across different graph sizes ($n = 10$ to 100). Box plots compare MTQA (isolated/non-isolated), PQA, QA, and SA. The red dashed line indicates the optimal solution obtained by CPLEX and Gurobi. All heuristic methods (quantum and classical SA) exhibit similar energy distributions close to the optimal ground state, indicating that for GPP, global parameterization (PQA) does not suffer the same catastrophic failure as in MVCP.}
\label{fig:fig6}
\end{figure*}

The analysis of the solution energy distributions provides the performance of different quantum annealing methods. Fig.~\ref{fig:fig6} presents energy distributions for graph sizes ranging from $n = 10$ to 100, comparing the performance of both MTQA approaches, PQA, QA, SA, and optimal CPLEX solutions.

Across all tested problem sizes ($n = 10$--100), all four quantum annealing approaches exhibited considerably similar energy distributions. The medians of the distributions for each method are consistently close to the optimal classical solutions, and no single quantum method demonstrates a consistent outperformance over the others in terms of the achieved solution energies for GPP. Although the distributions naturally broaden with increasing problem size, all methods maintain their primary concentrations of solutions near the optimal energy values.

\subsection{Eigenspectrum analysis}

We performed eigenspectrum\cite{Lanting2014A,SusaNishimori2020,SusaNishimori2021,Copenhaver2021,Crosson2014} analysis to investigate the quantum mechanical behavior of systems\cite{Dickson2013,Harris2018,PerdomoOrtiz2012,Boixo2013} under MTQA and PQA. This analysis focused on problems with 10 logical qubits constructed from the experimental problems used in our quantum annealing experiments. This problem size was selected as the largest feasible for complete eigenspectrum calculations within a 24-hour computational time limit.

\begin{figure}[b]
\centering
\includegraphics[width=\columnwidth]{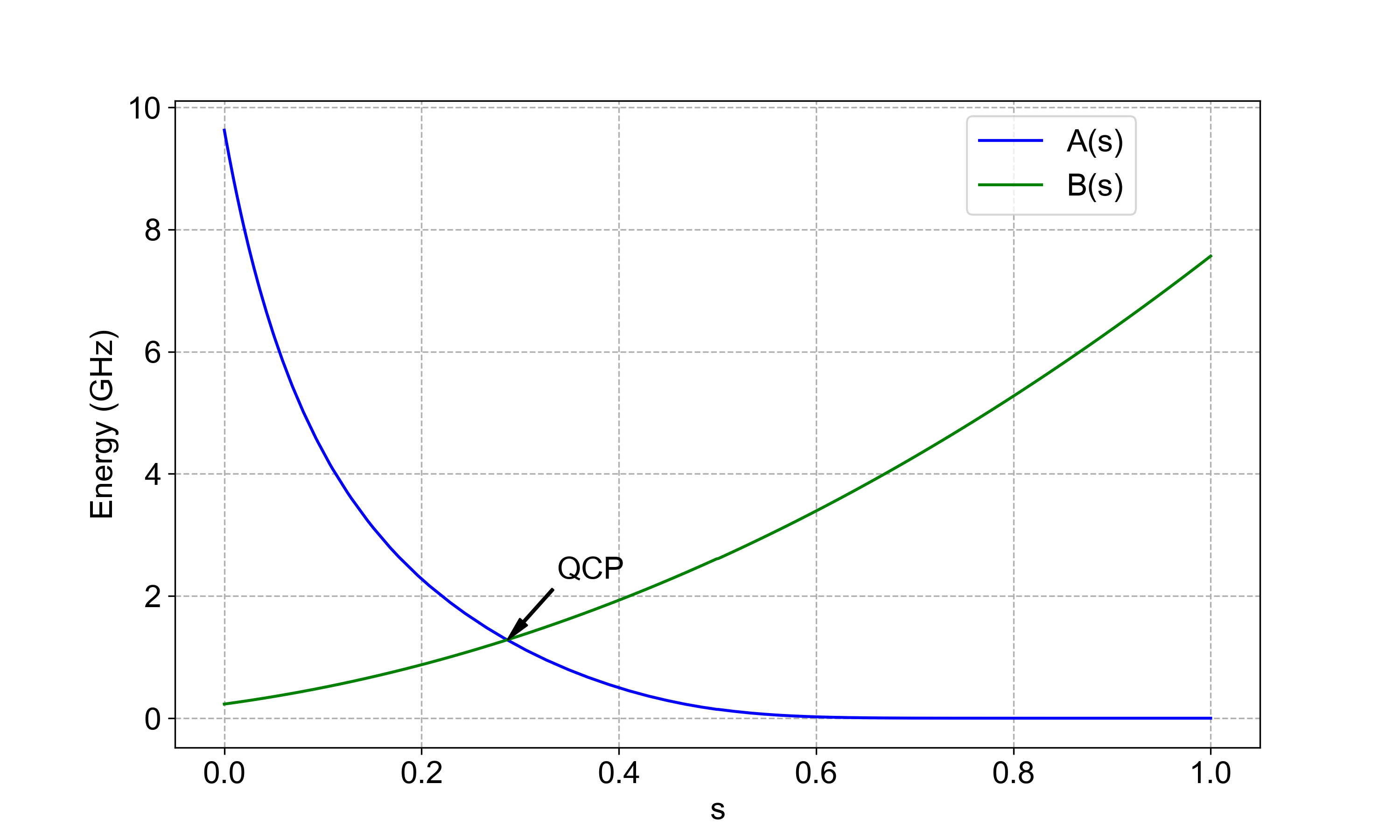}
\caption{D-Wave default annealing schedule showing the evolution of transverse field coefficient $A(s)$ and problem Hamiltonian coefficient $B(s)$ throughout normalized annealing time $s$. The quantum critical point (QCP) occurs at $s = 0.28$, where $A(s)=B(s)$ (approximately 1.28 GHz), marking the transition between quantum and classical regimes.}
\label{fig:fig7}
\end{figure}

Our analysis began with an examination of the D-Wave default annealing schedule\cite{DWaveQPU2024}, shown in Fig.~\ref{fig:fig7}, which illustrates the evolution of two critical energy terms: the transverse field coefficient $A(s)$ and the problem Hamiltonian coefficient $B(s)$. The annealing schedule demonstrates the controlled quantum evolution of the system, where $A(s)$ decreases from 9.62 GHz to zero, while $B(s)$ increases from 0.23 GHz to 7.56 GHz. A particularly significant feature occurs at $s = 0.28$, marked as the Quantum Critical Point (QCP). At this point, $A(s)=B(s)$ ($\approx 1.28$ GHz), representing a fundamental transition in the behavior of the system. Before the QCP, quantum fluctuations dominated by $A(s)$ drove the exploration of the solution space. After QCP, the problem Hamiltonian term $B(s)$ becomes dominant, guiding the system toward its final classical state.

To ensure that our analysis closely mirrors real-world quantum annealing conditions, we calculated eigenenergies using Hamiltonians derived from the embedded and scaled QUBOs. This approach incorporates the actual chain strengths used to maintain qubit coupling and the scaling factors applied to fit the hardware constraints, providing a more realistic representation of the behavior of the quantum system on physical hardware.

Using the linear algebra package (LAPACK)\cite{Harris2020NumPy,Anderson1999LAPACK}, we calculated the ground state and first excited state energies to analyze the evolution of the system state within its Hilbert space\cite{Halmos2012}, which is a complex vector space that provides a complete description of the system's quantum states\cite{Santhanam2009}. We examined three distinct scenarios: a single problem instance, two identical parallel problems, and two different parallel problems.

For a single problem (MVCP or GPP), we calculated the ground-state energy $E_{0,\mathrm{single}}$ and the first excited state energy $E_{1,\mathrm{single}}$, along with the energy gap, $\Delta_{\mathrm{single}} = E_{1,\mathrm{single}} - E_{0,\mathrm{single}}$. The results are presented in Figs.~8a and 8b, showing the distinct energy evolution patterns for each problem type.

When solving two identical instances of MVCP in parallel, the combined Hamiltonian was constructed in a block diagonal form, effectively doubling the Hamiltonian matrix dimensions without touching interactions between the two problem instances. The combined Hamiltonian can be expressed as follows:

\begin{equation}
H_{\mathrm{identical}} = \begin{bmatrix}
H_{\mathrm{MVCP}} & 0 \\
0 & H_{\mathrm{MVCP}}
\end{bmatrix}.
\end{equation}

This configuration doubles the ground-state energy $E_{0,\mathrm{parallel}} = 2E_{0,\mathrm{MVCP}}$ and adjusts the first excited state energy $E_{1,\mathrm{parallel}} = 2E_{0,\mathrm{MVCP}} + \Delta_{\mathrm{MVCP}}$ as shown in Fig.~8c. Thus, the energy gap remains the same as that for a single problem $\Delta_{\mathrm{parallel}} = \Delta_{\mathrm{MVCP}}$.

To solve the two different problems (MVCP and GPP), we employed a similar block diagonal structure:

\begin{equation}
H_{\mathrm{different}} = \begin{bmatrix}
H_{\mathrm{MVCP}} & 0 \\
0 & H_{\mathrm{GPP}}
\end{bmatrix}.
\end{equation}

This construction ensures that the Hamiltonian of each problem affects only its corresponding part of the system. As illustrated in Fig.~8d, the total ground-state energy becomes the sum of individual ground states $E_{0,\mathrm{combined}} = E_{0,\mathrm{MVCP}} + E_{0,\mathrm{GPP}}$, while the first excited state energy is $E_{1,\mathrm{combined}} = E_{0,\mathrm{MVCP}} + E_{0,\mathrm{GPP}} + \Delta_{\min}$. Where $\Delta_{\min}$ is the smaller of the individual energy gaps, $\Delta_{\min} = \min(\Delta_{\mathrm{MVCP}},\Delta_{\mathrm{GPP}})$.

This analysis indicates that MTQA can effectively handle multiple distinct problems, including three or more types, simultaneously. The block diagonal structure prevents level crossings between the ground and excited states, ensuring stability in quantum evolution. These findings provide theoretical support for scaling MTQA to more complex multi-problem implementations, reinforcing its potential for practical applications. An important implication of the block diagonal structure is that the minimum spectral gap of the combined system equals the smallest gap among the constituent problems. This has profound implications: the difficulty of successfully annealing the combined system is determined by the hardest constituent problem. In other words, adding easier problems to a batch does not increase the overall difficulty; however, the batch success probability cannot exceed that of the hardest individual problem. Because each problem occupies an independent subspace of the Hilbert space (enforced by the block diagonal structure), the local success probability for each problem instance is determined by its own spectral gap and not the combined gap. This means that MTQA can successfully solve easier problems in the batch even when harder problems fail, enabling partial batch success rather than all-or-nothing outcomes.

\begin{figure}[t]
\centering
\includegraphics[width=\columnwidth]{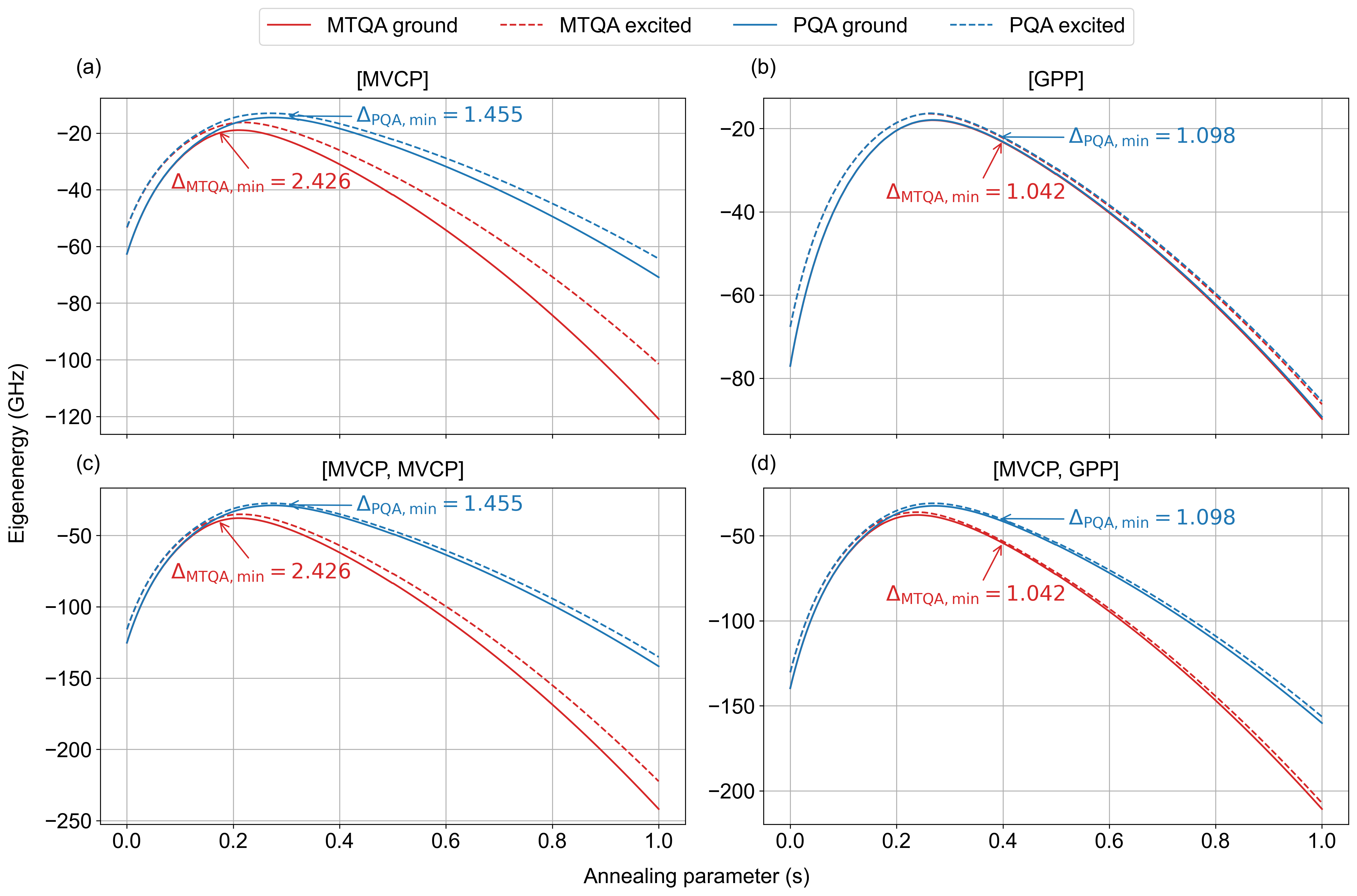}
\caption{Eigenspectrum analysis of parallel problem solving in quantum annealing using embedded Hamiltonians that incorporate actual chain strengths and scaling factors from hardware implementation. (a) Single MVCP problem eigenspectrum comparing MTQA (per-instance parameters) and PQA (global parameters) showing ground and first excited state energies. The global averaging of PQA significantly reduces the minimum energy gap for MVCP compared to MTQA. (b) Single GPP problem eigenspectrum comparison showing PQA results in slightly larger spectral gap than MTQA. (c) Two parallel MVCP problems demonstrating maintained energy gap consistency. (d) Combined MVCP-GPP system illustrating the additive nature of energy states with minimum gap determined by constituent problem with smallest gap. These results demonstrate that per-instance parameterization (MTQA) prevents spectral gap collapse that occurs with global parameterization (PQA) for MVCP.}
\label{fig:fig8}
\end{figure}

\subsection{Energy gap and state transition analysis}

\begin{figure*}[t]
\centering
\includegraphics[width=0.96\textwidth]{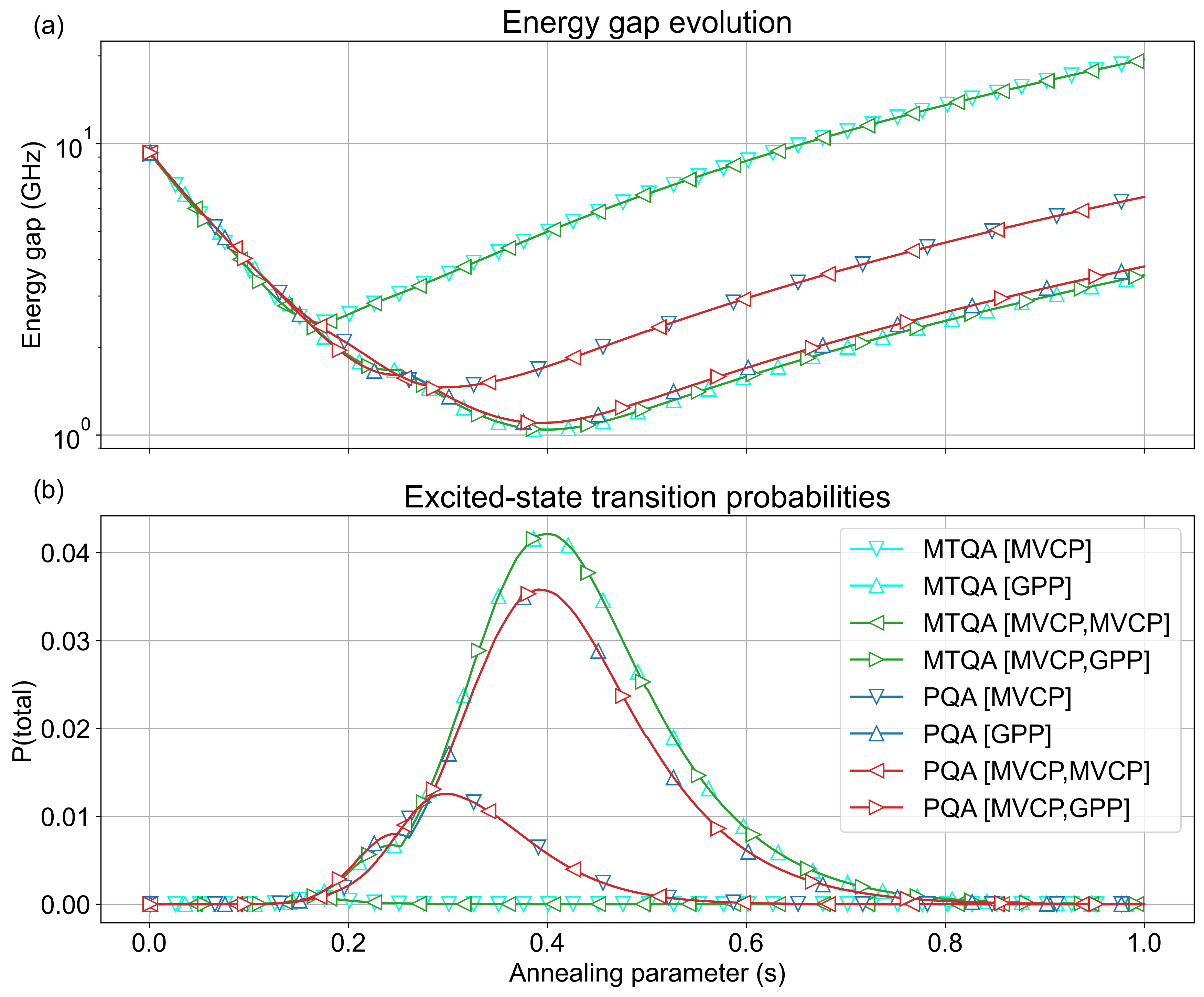}
\caption{Quantum mechanical analysis comparing MTQA, and PQA on embedded Hamiltonians. (a) Energy gap evolution throughout annealing process for all configurations. MTQA configurations maintain substantial energy gaps, while the global parameterization of PQA reduces gaps for MVCP. (b) Total excited state transition probabilities from Landau-Zener transitions and thermal excitations, demonstrating quantum stability with a maximum combined probability of 0.042 near $s \approx 0.41$.}
\label{fig:fig9}
\end{figure*}

The evolution of energy gaps during the annealing process provides crucial insights into system stability and transition dynamics. To examine this, we analyzed a closed quantum system, focusing solely on intrinsic quantum mechanical and thermal effects while excluding environmental noise. Figure~\ref{fig:fig9}(a) shows that all configurations maintain substantial energy gaps, with parallel implementations exhibiting patterns consistent with their constituent problems.

In our analysis, we examined two primary mechanisms influencing state: Landau-Zener (LZ) transitions\cite{Harris2018,Zener1932,Ivakhnenko2023,Amin2008,Albash2015,Lanting2014A} and thermal excitations\cite{Dickson2013,Ivakhnenko2023,Amin2008,Albash2015,Lanting2014A}. For the LZ transitions, we consider a two-level system described by the Hamiltonian\cite{Zener1932,Ivakhnenko2023}:

\begin{equation}
H(t) = -\frac{\Delta}{2}\sigma_x - \frac{\varepsilon(t)}{2}\sigma_z = -\frac{1}{2}
\begin{pmatrix}
\varepsilon & \Delta \\
\Delta & -\varepsilon
\end{pmatrix},
\end{equation}

where $\Delta$ is the minimum energy gap and $\varepsilon(t)=vt$ represents the linear time-dependent bias. The probability of a diabatic transition is given by the LZ formula:

\begin{equation}
P_{\mathrm{LZ}} = \exp(-2\pi\delta),
\end{equation}

where $\delta = \Delta^2/4\hbar v$ is the adiabaticity parameter, with $v=d\varepsilon/dt$ being the sweep velocity characterizing the rate of change of the energy bias\cite{Ivakhnenko2023}.

For thermal excitations, we employ the Boltzmann distribution with a complete partition function for our two-level system\cite{Lanting2014A}:

\begin{equation}
P_{\mathrm{thermal}} = \frac{w_1}{Z},
\end{equation}

where $Z = w_0 + w_1$ is the partition function, and $w_0 = \exp(-E_0/k_B T)$, $w_1 = \exp(-E_1/k_B T)$ are the Boltzmann factors for the ground and excited states respectively, with $T = 16$ mK is the system temperature (D-Wave Advantage 6.4). $E_0$ and $E_1$ are the energy levels of the ground and first excited states, and $k_B$ is the Boltzmann constant.

The total transition probability accounts for both mechanisms\cite{Amin2008}:

\begin{equation}
P_{\mathrm{total}} = P_{\mathrm{LZ}} + (1-P_{\mathrm{LZ}})P_{\mathrm{thermal}}.
\end{equation}

This formulation ensures thermal excitations are considered only if the system remains in the ground state after the LZ transition point. Our analysis, shown in Fig.~\ref{fig:fig9}(b), reveals that the maximum combined transition probability reaches 0.042 near $s \approx 0.41$ during the annealing process. Both parallel processing scenarios maintain transition probability profiles similar to their single-problem counterparts, with probability decreasing exponentially as the system evolves. By the end of annealing, total transition probabilities drop to near zero, indicating robust ground-state stability across all problem configurations.

This analysis demonstrates that combining multiple QUBOs into a single Hamiltonian through multi-tasking does not increase the computational complexity of the Hamiltonian. For two identical problems, the energy gap remained consistent with that of a single problem, indicating no additional challenges in maintaining the ground state during the annealing process. For two different problems, while the energy gap is influenced by the smaller of the individual gaps, the overall ground-state energy is simply the sum of the individual ground-state energies.

To investigate the physical origin of the PQA failure in MVCP, we compared the eigenspectrum of the embedded Hamiltonian ($n = 10$) under both MTQA (per-instance parameters) and PQA (global parameters). As shown in Fig.~\ref{fig:fig8}, the choice of parameterization dramatically alters the spectral gap. For MVCP, the global averaging used in PQA significantly reduced the minimum energy gap compared with MTQA. The minimum spectral gap determines the adiabatic condition for quantum annealing. A smaller gap increases the LZ transition probability, meaning that the system is more likely to undergo diabatic transitions to excited states rather than remaining in the ground state. For successful quantum annealing, the gap must be sufficiently large relative to the annealing speed to maintain adiabatic evolution. In Fig.~\ref{fig:fig8}, the closing of the gap increases the probability of diabatic transitions out of the ground state shown in Fig.~\ref{fig:fig9}; this explains the vanishing success probability observed for PQA in Figs.~\ref{fig:fig4} and \ref{fig:fig5}. However, as shown in Fig.~\ref{fig:fig8}(b), the global PQA settings resulted in a spectral gap that was slightly larger than that of MTQA. This incidental widening of the gap explains why PQA performed comparably to MTQA for GPP (Table~\ref{tab:gpp} and Fig.~\ref{fig:fig6}), despite the lack of local tuning. These results highlight that while global settings can occasionally suffice (as in GPP), they lack the reliability of MTQA's local approach, which prevents catastrophic gap closure in sensitive problems such as MVCP.

\section{Discussion}

Our investigation into MTQA reveals crucial insights into enhancing parallel quantum computation and resource optimization in quantum annealing systems. A key finding of this study is that MTQA processes multiple optimization problems simultaneously with solution quality comparable to traditional single-instance QA. It also significantly outperforms PQA, which utilizes global parameter settings, particularly for problem instances sensitive to specific annealing parameters.

The parallel embedding analysis shows that hardware utilization can be significantly improved through strategic problem placement. For smaller problems ($n < 30$), we achieved remarkable parallel processing capability with up to 130 embeddings. While this capacity decreases exponentially with problem size owing to increasing chain lengths and connectivity requirements, even larger problems ($n > 50$) maintain sufficient parallel embedding capacity to provide meaningful computational advantages.

Our eigenspectrum analysis of small-scale, idealized 10-logical-qubit systems provides theoretical support for the hypothesis that MTQA's approach of ensuring problem separability is consistent with preserving fundamental quantum coherence. Under the assumption of perfect problem isolation (modeled by a block-diagonal Hamiltonian), the analysis shows that energy gaps are maintained or determined by the constituent problem with the smallest gap. Thus, parallel embedding itself does not need to inherently increase the energetic complexity relevant to quantum evolution. This idealized theoretical finding aligns with MTQA's design philosophy of independent parameter control and is validated by our results, where MTQA, which employs such individualized control, successfully solved concurrent tasks while a PQA implementation with global parameters failed for certain problem types (e.g., MVCP).

The solution quality metrics further highlight the advantages of MTQA. For MVCP in Figs.~\ref{fig:fig4} and \ref{fig:fig5}, MTQA not only demonstrated GSP comparable to standard QA but also critically maintained this high performance, where our PQA implementation with global parameters showed a degradation (GSP dropping to zero for $n \ge 30$). Consequently, MTQA achieved a significantly reduced TTS compared to both QA and particularly the PQA (for which TTS became effectively infinite for $n \ge 30$). For GPP in Table~\ref{tab:gpp} and Fig.~\ref{fig:fig6}, MTQA, QA, SA, and PQA all produced good heuristic solutions with comparable median cut edges and solution energy distributions that were close to the classical optimum. In this context, the strength of MTQA lies in its robust framework that can deliver such GPP solutions effectively even when co-scheduling other potentially different tasks that might demand unique annealing parameters---a scenario where a global-parameter PQA could weaken, as seen with MVCP.

The theoretical framework developed for analyzing excited-state transitions in a closed system further supports the potential for quantum stability under the parallel processing conditions of MTQA. The analysis, combining Landau--Zener transitions and thermal effects, suggests that quantum coherence can be effectively maintained with exceptionally low total transition probabilities. This provides a theoretical rationale for the high GSP observed empirically with MTQA, even under multi-tasking. However, future studies should address the impact of environmental decoherence on real quantum devices. In practical hardware implementations, external noise and decoherence can affect the stability of solutions, and future work will explore noise-resilient strategies to mitigate these challenges.

Several specific application domains could benefit from the capabilities of MTQA: (1) Machine learning optimization: MTQA enables efficient multi-class classification by processing multiple binary classifiers simultaneously. For example, in multi-class Support Vector Machines (SVMs) using the One-vs-Rest strategy, MTQA can embed all binary classifiers in parallel, reducing the number of quantum annealing cycles from $C$ (number of classes) to a single cycle\cite{Artag2024MTQA}; (2) Decomposable optimization problems: Many real-world combinatorial optimization problems, such as the maximum clique problem, can be decomposed into smaller subproblems that are solved independently and then combined\cite{Pelofske2023Clique}. The ability of MTQA to solve multiple instances concurrently with per-instance parameter control makes it particularly suited for such decomposition strategies, where heterogeneous subproblems with different structural properties must be addressed simultaneously without compromising solution quality. (3) Multi-user cloud environments: As QPU access transitions to cloud-based services, MTQA provides a framework for efficiently handling concurrent user requests. Rather than sequentially processing small-to-medium scale problems that underutilize hardware capacity, MTQA can batch heterogeneous optimization tasks from multiple users, improving overall system throughput while maintaining solution quality through per-instance parameter control.

The ability of MTQA to reliably process multiple problems simultaneously---maintaining high solution quality where a PQA implementation with global parameters fails (as seen with MVCP) and performing comparably for other problems (GPP)---significantly enhances opportunities for optimizing quantum computing resources. This is relevant for efficiently handling batches of diverse or heterogeneous optimization tasks common in research and potential commercial applications. Furthermore, the MTQA framework, including optional isolation and per-instance parameter control, provides users with valuable flexibility and guidance for tailoring implementation choices to specific problem requirements and hardware characteristics.

However, certain limitations of the current study should be acknowledged to guide future research. It is important to note that the computational efficiency of MTQA is contingent upon sufficient hardware capacity. As the problem size increases, the chain lengths grow (Fig.~\ref{fig:fig3}) and the number of feasible parallel embeddings decreases exponentially (Fig.~\ref{fig:fig2}). For problems approaching the hardware capacity limits, the overhead of finding good minor embeddings can increase substantially, potentially diminishing the parallel processing advantage. Our results demonstrate clear benefits for problems up to $n = 100$ on D-Wave Advantage 6.4 hardware, but the optimal problem size range for MTQA deployment depends on the specific hardware topology and connectivity. Furthermore, our eigenspectrum and state transition analyses were based on small-scale idealized closed quantum systems. While providing valuable theoretical support, these models do not fully capture the impact of environmental noise, decoherence, and control errors present in real quantum devices. Finally, the pre-processing steps in MTQA, such as multiple embedding searches and individual parameter calculations, could introduce computational overhead, although this was not a bottleneck for the problem scales investigated.

\section{Methods}

We implemented MTQA on a D-Wave Advantage 6.4 quantum annealing processor, utilizing its Pegasus architecture with 5612 working qubits. All experiments were conducted using an annealing time of $20\,\mu\mathrm{s}$, the number of reads set to 2500 and the other parameters set to default. We accessed the D-Wave annealer through D-Wave Leap\cite{DWaveQPU2024}.

We generated test graphs using the Erd\H{o}s--R\'enyi\cite{Bollobas1998,Durrett2006,Hagberg2008} model with nodes ranging from 10 to 100 and an edge probability of 0.9. This model was selected for its ability to produce dense graphs, which pose significant challenges for optimization and provide a robust testbed for evaluating the effectiveness of MTQA against sequential QA methods.

\subsection{Minimum vertex cover problem (MVCP)}

MVCP aims to identify the smallest subset of vertices in a graph such that every edge has at least one endpoint in this subset. This problem is well-suited for formulation as a QUBO problem, making it solvable using quantum annealing techniques\cite{Karp2010,Lucas2014}. The QUBO formulation for MVCP is expressed as follows:

\begin{equation}
H_{\mathrm{MVCP}} = A\sum_{(i,j)\in E}(1-x_i)(1-x_j) + B\sum_{i\in V} x_i,
\end{equation}

where $x_i$ is a binary variable representing whether vertex $i$ is included in the vertex cover, $V$ is the set of vertices, and $E$ is the set of edges. Parameter $A$ serves as a penalty ensuring that all edges are covered, while $B$ penalizes the inclusion of each vertex, promoting a minimal vertex cover. To prioritize edge coverage over vertex inclusion, the parameters are chosen such that $0 < B < A$. Typically, $B = 1$ and $A = 2$ are used to effectively balance these objectives. This formulation ensures that minimizing $H_{\mathrm{MVCP}}$ corresponds directly to solving the MVCP\cite{Lucas2014}.

\subsection{Graph partitioning problem (GPP)}

GPP involves dividing the vertices of a graph into two subsets of equal size while minimizing the number of edges between these subsets\cite{Lucas2014,UshijimaMwesigwa2017,FuAnderson1986,MezardMontanari2009}. This problem can be effectively represented as a QUBO problem, making it suitable for quantum annealing. The QUBO formulation for GPP is expressed as follows:

\begin{equation}
    \begin{split}
        H_{\mathrm{GPP}} ={}& A\left(\sum_{i\in V} x_i - \frac{|V|}{2}\right)^2 \\
        & + B\sum_{(u,v)\in E} (x_u + x_v - 2x_u x_v),
    \end{split}
\end{equation}

where $x_i$ is a binary variable indicating the subset assignment of vertex $i$, and $x_u$, $x_v$ are binary variables associated with the edge between vertices $u$, $v$. The parameter $A$ enforces equal subset sizes, and $|V|$ is the total number of vertices in the graph. The formulation consists of two terms: the first term ensures that the subsets are of equal size, whereas the second term minimizes the number of edges between the subsets. To determine a lower bound for $A$, we set $B = 1$, and calculated as follows:

\begin{equation}
A \ge B\frac{\min(2\Delta,|V|)}{8},
\end{equation}

where $\Delta$ denotes the maximum degree of the graph. This choice balances the trade-off between penalizing unequal partition sizes and minimizing inter-subset edges\cite{Lucas2014}.

\subsection{MTQA implementation}

Our implementation of MTQA is structured into three main phases: parallel embedding search, quantum annealing execution, and solution extraction. The core of this approach is an iterative embedding strategy that efficiently maps multiple problems onto quantum hardware, as detailed in Algorithm~1.

\RestyleAlgo{ruled}
\SetKwComment{Comment}{/* }{ */}
\SetKwFunction{FindEmbedding}{FindEmbedding}
\SetKwFunction{RemoveNodes}{RemoveNodes}
\SetKwFunction{RemoveNodesAndNeighbors}{RemoveNodesAndNeighbors}
\SetKw{KwBreak}{break}
\SetKw{KwRet}{return}
\begin{algorithm}[hbt!]
\caption{Parallel Embedding Search}
\label{alg:parallel_embedding_search}
\KwData{Set of problems $P$, hardware graph $H$, isolation flag $I$}
\KwResult{\textit{embeddings} and \textit{order}}
\textit{embeddings} $\gets \{\}$\;
\textit{order} $\gets [\,]$\;
\While{true}{
    \textit{found} $\gets$ false\;
    \ForEach{$p \in P$}{
        \textit{embedding} $\gets$ \FindEmbedding{$p, H$}\;
        \If{\textit{embedding} exists}{
            \textit{embeddings}[$p$] $\gets$ \textit{embedding}\;
            append $p$ to \textit{order}\;
            \eIf{$I$}{
                \RemoveNodesAndNeighbors{$H$, \textit{embedding}}\;
            }{
                \RemoveNodes{$H$, \textit{embedding}}\;
            }
            \textit{found} $\gets$ true\;
        }
    }
    \If{\textit{found} = false}{
        \KwBreak\;
    }
}
\KwRet \textit{embeddings}, \textit{order}\;
\end{algorithm}

\subsection{Embedding strategy}

We utilized the minorminer tool\cite{Cai2014} from D-Wave's Ocean SDK to identify viable embeddings for each problem instance. This process ensures distinct mappings for parallel processing by allocating separate qubits to each problem. Additionally, our implementation incorporates an optional neighbor isolation strategy, which creates buffer zones between embedded problems. However, this comes at the cost of reduced total embedding capacity. For fair comparative analysis involving PQA, the PQA implementation used the identical physical qubit embeddings generated for the MTQA non-isolated runs to ensure a fair comparison of parameterization strategies.

\subsection{Chain strength calculation}

Appropriate chain strength ($J_{\mathrm{chain}}$) is critical in quantum annealing to ensure that logical qubits, represented by chains of physical qubits, act as single computational units throughout the annealing process\cite{Choi2008,Choi2011}. An insufficiently strong $J_{\mathrm{chain}}$ can lead to broken chains and invalid solutions, whereas an extremely strong $J_{\mathrm{chain}}$ can improperly flatten the energy landscape or introduce prohibitively small energy gaps, hindering the annealer's ability to find optimal solutions. For MTQA, which handles problems independently, we calculate the chain strength $J_{\mathrm{chain},\mathrm{MTQA}_p}$ for each problem instance $p$ using problem-type-specific methods and prefactors, ensuring tailored coupling for diverse parallel workloads.

To address the distinct characteristics of our chosen problem types, MVCP and GPP, we employed two established heuristic methods for determining the base chain strength, subsequently modified by empirically optimized prefactors ($\alpha_p$).

\subsection{Uniform torque compensation}

Given that MVCP instances often exhibit relatively uniform interaction strengths, we utilized a chain strength calculation based on the \texttt{uniform\_torque\_compensation} functionality provided by D-Wave\cite{DWaveQPU2024}, which adjusts chain couplings to maintain balanced magnetic torques across qubits. This method balances the magnetic torque on qubit chains and is expressed as:

\begin{equation}
J_{\mathrm{base},\mathrm{MVCP}} = \sqrt{\frac{\sum J_{ij}^{2}}{N_J}}\sqrt{\bar d},
\end{equation}

where $\sum J_{ij}^{2}/N_J$ is the mean square of the problem's quadratic biases (coupler strengths in the logical QUBO), and $\bar d$ is the average degree of the logical problem graph. For our MTQA implementation, we applied an empirically optimized overall prefactor $\alpha_{\mathrm{MVCP}} = 0.5$:

\begin{equation}
J_{\mathrm{chain},\mathrm{MTQA}_{\mathrm{MVCP}}} = \alpha_{\mathrm{MVCP}}J_{\mathrm{base},\mathrm{MVCP}},
\end{equation}

This lower effective prefactor allowed sufficient quantum fluctuations for effective exploration of the MVCP solution space while maintaining chain integrity.

\subsection{Scaled chain strength (GPP)}

GPP instances often feature more complex connectivity and a wider range of interaction strengths. For these, we utilized a ``scaled'' functionality provided by D-Wave\cite{DWaveQPU2024} for the base chain strength, which sets the strength relative to the maximum absolute bias in the problem:

\begin{equation}
J_{\mathrm{base},\mathrm{GPP}} = \max(|h_i|,|J_{ij}|),
\end{equation}

where $\max(|h_i|,|J_{ij}|)$ is the maximum absolute value among all linear biases ($h_i$) and quadratic biases ($J_{ij}$) in the logical QUBO. We then applied an empirically optimized overall prefactor $\alpha_{\mathrm{GPP}} = 1.5$:

\begin{equation}
J_{\mathrm{chain},\mathrm{MTQA}_{\mathrm{GPP}}} = \alpha_{\mathrm{GPP}}J_{\mathrm{base},\mathrm{GPP}},
\end{equation}

This stronger effective coupling ensures chain integrity even under the competing forces inherent in graph-partitioning constraints. The final prefactors, $\alpha_{\mathrm{MVCP}} = 0.5$ and $\alpha_{\mathrm{GPP}} = 1.5$, were selected after systematic empirical evaluation across a range of values (0.5 to 2.0), choosing those that consistently yielded the best balance of high GSP and minimized TTS. This approach robustly maintains chain coherence without over-constraining the system, which could inhibit beneficial quantum tunneling effects.

\subsection{QUBO scaling}

To fit within the hardware constraints of the D-Wave quantum annealer, each embedded QUBO was individually scaled. Our scaling methodology examines both individual qubit biases ($h_i$) and pairwise coupling strengths ($J_{ij}$) within each problem instance to calculate a scaling factor for that instance as defined by D-Wave\cite{DWaveQPU2024}, ensuring all terms remain within the hardware's operational ranges. The approach considers the maximum values of the diagonal and off-diagonal terms, as well as the total interaction strength per component. By scaling each embedded QUBO independently, smaller problem instances are not disproportionately affected by the scaling requirements of larger ones during simultaneous problem solving. This approach preserves the mathematical structure of each problem while adhering to hardware limitations, ensuring accurate representation of the original optimization problems on the quantum hardware.

\subsection{Unembedding strategy}

The translation of physical qubit states back to logical variables\cite{Choi2008,Choi2011,Vinci2015} employed the majority vote unembedding method from D-Wave's Ocean SDK. While alternative approaches, such as weighted random or energy minimization, could improve solution precision, and MTQA's framework supports the application of such diverse strategies on a per-instance basis within parallel problems, we chose majority voting to demonstrate MTQA's fundamental effectiveness in parallel problem solving. This approach provides a clear and reliable baseline for evaluating the multitasking capabilities of quantum annealing systems while maintaining solution consistency.

\subsection{PQA implementation for comparison}

To directly compare MTQA with PQA that employs more global parameter settings, we implemented a comparative PQA setup. This PQA implementation processed the same problem instances (MVCP or GPP) in parallel as MTQA. For a fair comparison, particularly with MTQA non-isolated, the PQA runs for a given set of parallel instances utilized the identical physical qubit embeddings that were generated and used by the MTQA non-isolated configuration. Unlike the per-instance scaling of MTQA, for this PQA implementation, the QUBOs for parallel problems were scaled collectively using a single global scaling factor. This factor was determined by considering the maximum coefficient magnitudes across all embedded QUBOs in parallel to ensure all fit within hardware limits. Then, a global approach to chain strength was used. Specifically, for parallel problems (e.g., MVCP and GPP), the chain strength $J_{\mathrm{chain},\mathrm{PQA}}$ was set to the average of the individual $J_{\mathrm{chain},\mathrm{MTQA}_p}$ values that MTQA would have used for those same instances.

A global majority vote was applied to all problem instances within PQA as an unembedding method. This PQA implementation was designed to reflect a scenario in which individual problem characteristics are less granularly addressed in parameter settings compared to MTQA.

\subsection{Simulated Annealing}

For the classical heuristic comparison, we employed SA using the D-Wave Ocean SDK (\texttt{neal.SimulatedAnnealingSampler})\cite{DWaveQPU2024}. The annealing schedule was set to default parameters with 1 000 sweeps per read and 2 500 reads per instance, ensuring a fair baseline for comparing ground-state probabilities and solution energies against quantum approaches.

\subsection{Classical solvers}

To benchmark the performance of MTQA, we utilized CPLEX and Gurobi as classical optimization solvers. Both solvers translated the QUBO formulations into binary optimization problems using a shared framework. CPLEX leveraged its native binary quadratic programming capabilities, whereas Gurobi utilized its built-in mixed-integer programming (MIP) framework. Both solvers were configured with a 600-second time limit per instance, and the MIP Focus parameter was set to 2, prioritizing solution optimality over the rapid discovery of feasible solutions.

For MVCP instances, both solvers achieved optimal solutions within the specified time limit for all problem sizes, establishing reliable baselines for MTQA performance. However, for GPP instances, the performance varied with the problem size. CPLEX reached the time limit for problems with 60 or more nodes, whereas Gurobi encountered time limits starting at 30 nodes. Despite these constraints, both solvers provided valuable metrics for comparing the MTQA efficiency, particularly in terms of solution quality and computation time across a range of problem sizes.

\section*{List of abbreviations}

\begin{description}
\item[DC-QAOA] Digitized-Counterdiabatic Quantum Approximate Optimization Algorithm
\item[GPP] Graph partitioning problem
\item[GSP] Ground-state probability
\item[ICE] Integrated control error
\item[LAPACK] Linear Algebra Package
\item[LZ] Landau-Zener
\item[MIP] Mixed-integer programming
\item[MTQA] Multi-tasking quantum annealing
\item[MVCP] Minimum vertex cover problem
\item[NP] Nondeterministic Polynomial
\item[PQA] Parallel Quantum Annealing
\item[QA] Quantum Annealing
\item[QCP] Quantum Critical Point
\item[QPU] Quantum Processing Unit
\item[QUBO] Quadratic unconstrained binary optimization
\item[SA] Simulated annealing
\item[SVM] Support Vector Machines
\item[TTS] Time-to-solution
\end{description}

\section*{Declarations}

\subsection*{Availability of data and materials}
All data and code generated and/or analyzed during this study are available at:
\url{https://github.com/shrakashlab/MTQA}

\subsection*{Competing interests}
The authors declare that they have no competing interests.

\subsection*{Funding}
Not applicable

\subsection*{Author contributions}
J.A. conceptualized the study, developed the methodology, conducted the experiments, analyzed the data, and drafted the manuscript. K.A. and T.K. provided technical guidance on experimental design, conducted the experiments, analyzed the data, and manuscript preparation. D.T. and M.S. provided technical guidance on experimental design and manuscript preparation. J.S. supervised the research, provided critical feedback on the methodology, and contributed to the interpretation of the results. All authors reviewed and approved the final manuscript.

\begin{acknowledgments}
Not applicable
\end{acknowledgments}

\bibliography{reference}

\end{document}